\documentclass[aps,physrev,twocolumn,unsortedaddress,superscriptaddress,showkeys,nofootinbib,10pt]{revtex4-2}

\usepackage{graphicx}
\usepackage{dcolumn}
\usepackage{bm}
\usepackage[dvipsnames]{xcolor}
\usepackage[colorlinks=true,urlcolor=Blue,linkcolor=NavyBlue,citecolor=NavyBlue]{hyperref}
\usepackage{amsfonts}
\usepackage{empheq}
\usepackage{amsmath}
\usepackage{amssymb}

\newcommand{\MBH}{M_{{\rm BH}}}
\newcommand{\Msun}{M_\odot}
\newcommand{\Rsch}{R_{{\rm sch}}}
\newcommand{\Rinf}{R_{{\rm inf}}}
\newcommand{\Egr}{\mathcal{E}}
\newcommand{\Veff}{\mathcal{V}_{{\rm eff}}}
\newcommand{\mth}{m_{{\rm th}}}

\graphicspath{{./}{Plots/}}

\begin{document}
\title{Fermionic Dark Matter Spikes: origin and growth of black hole seeds}
\author{Valentina Crespi} 
\email[]{Contact author: valentinacrespi@fcaglp.unlp.edu.ar}
\affiliation{Instituto de Astrofísica de La Plata, UNLP-CONICET, Paseo del Bosque s/n B1900FWA La Plata, Argentina}
\affiliation{Facultad de Ciencias Astronómicas y Geofísicas, UNLP, Paseo del Bosque s/n B1900FWA La Plata, Argentina}

\author{Carlos R. Argüelles}
\email[]{carguelles@fcaglp.unlp.edu.ar}
\affiliation{Instituto de Astrofísica de La Plata, UNLP-CONICET, Paseo del Bosque s/n B1900FWA La Plata, Argentina}
\affiliation{Facultad de Ciencias Astronómicas y Geofísicas, UNLP, Paseo del Bosque s/n B1900FWA La Plata, Argentina}
\affiliation{ICRANet, Piazza della Repubblica 10, I-65122 Pescara, Italy}

\author{Jorge A. Rueda}
\email[]{jorge.rueda@icra.it}
\affiliation{ICRANet, Piazza della Repubblica 10, I-65122 Pescara, Italy}
\affiliation{ICRANet-Ferrara, Dip. di Fisica e Scienze della Terra, Universit\`a di Ferrara, Via Saragat 1, I--44122 Ferrara, Italy}
\affiliation{ICRA, Dipartamento di Fisica, Sapienza Universit\`a  di Roma, Piazzale Aldo Moro 5, I-00185 Rome, Italy}
\affiliation{Dipartimento di Fisica e Scienze della Terra, 
Universit\`a di Ferrara, Via Saragat 1, I-44122 Ferrara, Italy}
\affiliation{INAF, Istituto di Astrofisica e Planetologia Spaziali, Via Fosso del Cavaliere 100, 00133 Rome, Italy}

\date{\today}

\begin{abstract}
    We characterize the overdensity (\textit{spike}) of fermionic dark matter (DM) particles around a supermassive black hole (SMBH) within a general relativistic analysis. The initial DM halo distribution is obtained by solving the equilibrium equations of a self-gravitating system of massive fermions at a finite temperature, according to the Ruffini-Arg\"uelles-Rueda (RAR) model. The final fermionic DM spike is calculated around a Schwarzschild SMBH. We explore two possible interpretations for the origin and evolution of the SMBH seed. One corresponds to the traditional scenario, where a small BH mass of ordinary baryonic origin sits at the halo's center and grows adiabatically. The other scenario is of DM origin, where the dense and degenerate fermion core predicted by the RAR model grows adiabatically by capturing baryons until its gravitational collapse, becoming a heavy SMBH, whose specific value depends on the fermion mass. We study various initial fermionic DM profiles that the theory allows. We show that overall dilute (i.e., Boltzmannian) fermionic DM develops the well-known spike with density profile $\rho \sim r^{-3/2}$. Instead, for semi-degenerate fermions with a dense and compact core surrounded by a diluted halo, we find novel spike profiles that depend on the particle mass and nature. In the more general case, fermionic spikes do not develop a simple power-law profile. Furthermore, in some cases, the SMBH depletes the surrounding DM density instead of enhancing it. 
    Thus, the self-consistent inclusion of the DM candidate nature and mass in determining the structure and distribution of DM in galaxies, including the DM spikes around SMBHs, is essential for the specification of DM astrophysical probes such as BH mergers, gravitational waves, or stellar orbits.
\end{abstract}


\maketitle

\section{Introduction} \label{introduction}
In recent decades, it has been well established that dark matter (DM) particles may conglomerate around a central black hole (BH), forming an overdensity in the local surroundings \citep{gondolo1999dark,ullio2001dark,merritt2004evolution,sadeghian2013dark}. The first heuristic calculations were aimed at baryonic matter cusps by computing the density of a star cluster near a collapsed object, assuming a distribution function given by a power law in the star's energy \citep{1972ApJ...178..371P}. Later on, \citet{young1980numerical} developed a numerical model based on radial action conservation of the stars, recovering Peebles' main result. There, in the center of the embedding distribution ($\rho_i \sim r^{-2}$), a density cusp was obtained with $\rho_f \sim r^{-3/2}$. 
This work was later extended by \cite{1995ApJ...440..554Q} with the adoption of the isochrone model \citep{henon1960evasion} as an example of models with analytic cores, different from the isothermal law, and $\gamma$-models \citep{dehnen1993family} for non-analytic cores. They derived a relation for the initial and final slope of the density distribution by scaling relations of the particle's energy and radii and imposing the conservation of the total mass and angular momentum. 

\citet{gondolo1999dark} (hereafter G\&S) applied these seminal ideas to DM halos with initial power-law cusps as suggested by N-body simulations. They explored different initial power-law DM distributions $\rho_i(r)\propto r^{-\gamma}$ with $0<\gamma<2$, coining for the resulting central over-densities $\rho_f(r)\propto r^{-\gamma_{sp}}$ with $2.25<\gamma_{sp}<2.5$, the name of DM spikes. Like their predecessors, the resulting DM density spike was calculated in the Newtonian regime. However, G\&S assumed \textit{ad-hoc} integration limits to incorporate the effects of gravitational capture of the central BH. This resulted in an inner (starting) radius for the DM density, given by $4\Rsch$, where $\Rsch$ is the Schwarzschild radius. \citet{sadeghian2013dark} improved such an approximation by using a general relativistic treatment, leading to an innermost boundary radius of $2\Rsch$, with consequent higher densities for the DM spike.

Several studies have suggested that DM overdensities at galactic centers may be attenuated by dynamical processes and different initial conditions for BH growth. Some examples are off-centered BH seeds, mergers of host galaxies, and baryon adiabatic contraction \citep{ullio2001dark,merritt2002dark,merritt2004evolution,bertone2005time}. More generally, open problems such as the origin, location, and mass of the initial BH seed and the consequent timescale needed to reach a given spike have recently started to be studied within specific BH-formation scenarios in a cosmological framework (see, e.g., \cite{2024arXiv240408731B}).

Thus, our aim here is to obtain a more realistic description of DM spikes than obtained in G\&S by considering some of these open issues for fermionic DM profiles within a fully relativistic treatment, providing key insights into the nature and mass of the DM particle and possible origin and growth of the initial BH-seeds. The need for a more realistic treatment of galaxy centers is justified by indirect DM searches, gravitational waves (GWs) detection, and a better understanding of the motion of surrounding stars. 

In particular, the center of the Milky Way has been one of the most attractive targets for DM indirect searches.
If DM particles self-interact, these enhancements should lead to potentially detectable fluxes of high-energy radiation emanating from the central regions \citep{belikov2014diffuse,lacroix2014probing,yang2024searching}. 
Many projects and experiments have been developed to constrain the DM annihilation and decay rates and to model the emission coming from the galactic center \citep{abramowski2011search,gomez2013constraints,daylan2016characterization,2020PDU....3000699Y}.
Likewise, observations of S-cluster stars around SgrA* were used to put constraints on the size and mass of the spike \citep{2018A&A...619A..46L,Heissel:2021pcw,GRAVITY:2021xju,GRAVITY:2023cjt,shen2024exploring}.
DM spikes could also be a relevant target for GW detectors \citep{bertone2018new}. In principle, detecting and characterizing DM spikes around BHs is feasible by measuring their influence on the gravitational waveform during BH mergers  \citep{eda2015gravitational,yue2019dark,cardoso2020constraints,kavanagh2020detecting}. However, observationally, there is no definitive evidence either in favor or against the existence of DM density enhancements at the centers of galaxies. 

At this point, it is worth recalling that the self-consistent inclusion of the DM nature and mass in the determination of the behavior, structure, and distribution of DM in galaxies, including the DM spikes around SMBHs, is essential for the specification of DM astrophysical probes or the analysis of observational data. This paper aims to characterize the DM spikes around SMBHs if the DM particle is a massive, neutral fermion. Bosonic matter fields (either bare or self-interacting) are also possible DM candidates which are extensively explored in the literature  (see e.g. \cite{2024FrASS..1147518M} for a brief review). 
Considerable attention is being dedicated to the case of ultralight bosons with $m\sim 10^{-22}$ eV, since they naturally arise in string theory models while at the same time the predicted DM density abundance is comparable with the present DM energy density of the Universe \citep{2016PhR...643....1M,2017PhRvD..95d3541H}. 
However, even if such ultralight DM candidates can explain the large scale structure similarly as in $\Lambda$CDM models (see e.g. \citep{2021MNRAS.506.2603M}), strong tension exist on small scales, e.g., they cannot explain the Lyman-$\alpha$ forest and the dwarf galaxies core sizes simultaneously\footnote{It is expected that by adding new degrees of freedom to standard ultralight DM in the form of specific self-interacting terms may release those tensions \citep{2023MNRAS.521.2608M}, though only preliminary results have been shown so far.} \cite{2017PhRvL.119c1302I}. 
Instead, fermionic DM candidates in the mass range of $\mathcal{O}(10^1$--$10^2)$ keV under the RAR model do not suffer from any of those problems, as explained in \citep{2019PDU....24..278A,arguelles2023fermionic}.

In this work, we apply the general relativistic approach derived in \cite{sadeghian2013dark} and adopt as the initial DM halo the equilibrium configurations of self-gravitating fermions, obtained with the extended Ruffini-Arg\"uelles-Rueda (RAR) model \citep{2015MNRAS.451..622R,2018PDU....21...82A}. 
Our motivation to work within such a fermionic DM context is (at least) twofold. First, the massive $\mathcal{O}(10^1$--$10^2)$ keV fermions conforming the DM halos are in agreement with the large-scale structure of the Universe and DM halo formation: in \cite{2021MNRAS.502.4227A}, it was demonstrated that in a non-linear structure formation theory based on a Maximum Entropy Production Principle for fermions, the most general solution is stable and develops a degenerate compact core surrounded by a diluted halo. 
Such a core-halo density profile explains a variety of galactic observables. In \citep{2023ApJ...945....1K} the authors performed an analysis to fit $\sim$ 120 galactic rotation curves using recent SPARC data, modeling DM halos with fermionic core-halo profiles and compared with other phenomenological models.
In \cite{2024A&A...689A.194M} the phase-space data of the GD-1 stellar stream is explained with fermionic core-halo model whilst the core explains the dynamics of the S star cluster at $\sim$ mpc scales.
An extensive analysis on the regimes of these fermionic models and associated morphologies was done in \cite{2019PDU....24..278A,2023ApJ...945....1K} with a study of galactic universal relations. Recently, a stability analysis devoted to the Milky Way was performed in \cite{krut2025thermodynamics}. The application of the extended RAR model to the innermost S-cluster stars near the Milky Way's center without assuming a BH has been presented in \citep{2020A&A...641A..34B,2021MNRAS.505L..64B}, including relativistic orbital precession \cite{2022MNRAS.511L..35A}.
Thus, it provides a novel DM profile that resembles those obtained in N-body simulations in the outer halo \citep{2023ApJ...945....1K}, but that differ toward the center, where they develop a highly dense core. The latter has never been analyzed in the context of DM spikes once a central BH seed is formed.

Second, the presence of the dense DM core has an additional astrophysical and cosmological appeal: under given conditions and for a given particle mass, it becomes unstable against gravitational collapse, forming a SMBH at a well-established critical mass which is still surrounded by the halo \citep{2021MNRAS.502.4227A,2023MNRAS.523.2209A,2024ApJ...961L..10A}. Thus, this fermionic theory provides a novel mechanism for the formation of SMBHs, with key applications to the open problem of origin and SMBH seeds in the high-redshift Universe \cite{2024ApJ...961L..10A}. For instance, the collapse of a core of $m = 100$ keV fermions form a SMBH of $M_{\rm BH} = 6.3\times 10^7 M_\odot$. The heavier the fermion, the lighter the SMBH. Starting from these heavy BH seeds, SMBHs of a few $10^9 M_\odot$ form in less than a Gyr by sub-Eddington accretion \cite{2023MNRAS.523.2209A}. Recently, within the context of ultralight DM with attractive self-interactions, it was shown that bosonic halos can form dense solitons at their center which may reach a critical mass of collapse into a BH \citep{2024MNRAS.533.2454P}, though only preliminary Newtonian simulations with limited numerical resolution have been so far obtained.

Further, \citet{2024ApJ...961L..10A} showed that the sedimentation of ordinary matter (e.g., by accretion or galaxy mergers) at the bottom of the dense fermionic DM core triggers its gravitational collapse at a critical mass value up to $\approx 40\%$ lower than the critical mass of a pure DM core. The above scenario strongly motivates exploring the consequences of this SMBH formation channel at the cosmological and galactic levels. Indeed, concerning the high-redshift Universe, as shown in \cite{2024ApJ...961L..10A}, it could explain the unexpectedly observed population of SMBHs in the early Universe by the James Webb Space Telescope \citep{2022A&A...666A..17G,2023ApJ...954L...4K,2023ARA&A..61..373F}. 

One of the main goals of this paper is to assess how the fermionic DM density profile redistributes due to a newly formed SMBH at its center. We will consider in Sec. \ref{sec:Results} two different scenarios of the nature, origin, and evolution of the central BH: the G\&S scenario of a small BH of ordinary baryonic origin and the SMBH seed from the collapse of a dense DM fermion core.

The outline of this work consists of describing the theoretical framework to obtain the fermionic DM mass density in section \ref{sec:the method}, together with a description of the fermionic DM model within the RAR theory, including its relevance in the process of formation of central SMBHs recently shown in \citet{2024ApJ...961L..10A}. 
Section \ref{sec:Results} presents the results for several fermionic DM overdensities, depending on the nature of the BH seed considered, and compares the obtained spikes with previous ones in the literature. 
In section \ref{sec:conclusions}, we summarize the results of this work, including the current limitations of the theoretical framework and possible extensions, as well as potential applications to be further analyzed in forthcoming research. Appendix \ref{appendix-A} details the use of adiabatic invariants to relate particle energies, and Appendix \ref{appendix-B} develops on the calculation to obtain the distribution of DM when a SMBH suddenly appears in the center.

\section{Theoretical Framework} \label{sec:the method}

\subsection{Relativistic framework} \label{subsec:framework}

We consider the seminal work by G\&S for the calculation of the particle density profile after a central SMBH is adiabatically formed in the center of a DM halo. Specifically, we follow the fully relativistic treatment of this procedure generalized by \cite[sec.III]{sadeghian2013dark} for the growth of a Schwarzschild BH. 

For a statistical description of an array of particles in a given spacetime, we can write the mass-current four-vector as a first moment (in $p^\mu /m$) of the distribution function (DF) $f(\mathbf{x},\mathbf{p})$ \citep[Ch.IV]{Synge1960-SYNRTG}
\begin{equation}
    J^\mu(\mathbf{x}) = \int f(\mathbf{x},\mathbf{p}) \frac{p^\mu}{m}\sqrt{-g}\ d^4p,
\end{equation}
where $m$ is the particle rest mass, $p^\mu$ the four momentum and $g$ is the determinant of the underlying metric. We hereafter adopt geometric units $c=G=1$ and restore astrophysical units in the quantitative examples.

In this work, we will consider the re-distribution of DM particles -- its orbits -- around a Schwarzschild SMBH once it has adiabatically grown. The metric of the spherically symmetric Schwarzschild spacetime for a BH of mass $\MBH$ is
\begin{subequations}
    \begin{align}\label{eq:ds2}
    &ds^2 = g_{00}(r) dt^2 +  g_{11}(r) dr^2 - r^2 d\theta^2 - r^2 \sin\theta d\phi^2,\\
    &g_{00}(r) = - \frac{1}{g_{11}(r)} = \left(1-\frac{2\MBH}{r}\right). \label{eq:g00}
\end{align}
\end{subequations}

To account for BH capture effects, it is convenient to introduce the constants of motion, specifying the dependence on the particle's angular momentum, $L$. Any particle with $L < L_{\rm cr} = 2\sqrt{3}\ \MBH$ will cross the BH horizon. The mass density, related to the mass current by $\rho(r)=\sqrt{g_{00}(r)} J^0$, becomes
\begin{equation} \label{eq:rho}     
    \rho(r) = \frac{4\pi m^4}{r^2 \sqrt{g_{00}}} \int_{\Egr_{1}}^{\Egr_{2}} \Egr d\Egr \int_{L_{1}}^{L_{2}}\frac{f(\Egr,L)\ L\ dL}{\sqrt{\Egr^2 - g_{00}(1+\frac{L^2}{r^2})}},
\end{equation}
where $\Egr$ and $L$ are the particle conserved (along the geodesic) energy and angular momentum per unit rest mass. Because these integrals are computed for every radial position $r$, the lower and upper limits must be specified at every position. For this task, we now analyze the condition to have bound orbits. With the aid of the radial motion effective potential, $\Veff^2(r,L)=g_{00}(r)(1+L^2/r^2)$ with $g_{00}$ given by Eq. (\ref{eq:g00}), we find the following limits
\begin{align}
    (L_1)^2 &= \frac{27\Egr^4\ -\ 36\Egr^2\ +\ 8\ +\ \Egr (9\Egr^2-8)^{3/2}}{2(\Egr-1)}, \label{eq:Lmin} \\
   (L_2)^2 &= r\left(\frac{\Egr^2}{g_{00}(r)} -1\right), \label{eq:Lmax} \\
   \Egr_1 &= \begin{cases}
       (1-2/r)/\sqrt{1-3/r}, &  4 \leq r/\MBH \leq 6,\\
       (1+2/r)/\sqrt{1+6/r}, & r/\MBH \geq 6, 
       \end{cases}\label{eq:Emin} \\
    \Egr_2 &= 1 .
\end{align}
Equation (\ref{eq:Lmin}) corresponds to the maximum of $\Veff$ achieved when $\Veff(r_{1},L_{1})=\Egr$ ($r_1$ is the location of the maximum, i.e., $d\Veff/dr=0$, $d^2\Veff/dr^2<0$) and is equivalent to Eq. (3.16) 
in \cite{sadeghian2013dark}. Equation (\ref{eq:Lmax}) corresponds to the inner turning point $\Veff(r,L)=\Egr$ ($\dot{r}=0$), and Eq. (\ref{eq:Emin}) to energies equal to the maximum of $\Veff$, $\Veff(r,L_{1}(\Egr_1))=\Egr_1$, and solving for $\Egr_1$.

The DF in eq. (\ref{eq:rho}), $f(\Egr,L)$, which describes the distribution of DM particles under the influence of the grown BH, is \textit{a priori} unknown. However, under the assumption of adiabatic growth of the initial BH seed, the final form of the DF can be related to the known initial DF for particles orbiting in a self-gravitating DM halo. This approach is valid when the changes in the BH gravitational potential occur on slow timescales compared to dynamical periods of particles in regions where the BH dominates. The BH gravitational influence radius can be estimated as $\Rinf = c^2\Rsch/2\sigma^2 \sim 10^6 \Rsch$, where $\sigma$ is a typical velocity dispersion dictated by the BH gravitational potential \citep[Ch. 3.6]{binney2011galactic}. Indeed, this is the case for central orbits in a galaxy with periods $t_{{\rm dyn}} \lesssim \Rinf/ \sigma \sim 10^4$ yr. In turn, the SMBH growth timescale is set by the accretion timescale, $t_{{\rm BH}}=\MBH/\dot{M}$. If we adopt accretion rates lower than the Eddington limit, the latter provides a lower limit to the BH growth timescale, $t_{{\rm BH}}\sim 10^7$ yr.

Thus, throughout this adiabatic growth, the action variables of the particles remain constant, allowing us to relate the initial and final states in energy and angular momentum of the DM orbits. In Appendix \ref{appendix-A}, we present the numerical computation of this relation.
We apply this formalism in: \textsc{i}) a negligible BH-seed adiabatically growing in the center of a fermionic RAR halo, namely, Model I  see section \ref{sec: model 1}), and \textsc{ii}) a DM fermion-core accreting baryons and, hence, adiabatically growing to its critical state for gravitational collapse, corresponding to Model II (as detailed in section \ref{sec: model 2}).

\subsection{Fermionic DM halos: extended RAR model} \label{subsec:RAR model}
The model describes self-gravitating halos in hydrostatic and thermodynamic equilibrium in a full general relativistic framework. It consists of neutral, massive fermions with spin $1/2$, with its equation of state accounting for tidal truncation of particles with large momenta. 
These equilibrium configurations can be derived via a maximum entropy production (MEP) principle, where the total entropy $S_{\rm tot}$ maximizes at fixed total particle number $N_{\rm tot}$ and total energy $E_{\rm tot}$ \citep{chavanis2015models,2020EPJP..135..290C,krut2025thermodynamics}
\begin{equation}
    \text{max} \left\{ S_{\rm tot}\ |\ N_{\rm tot},E_{\rm tot}\ \text{fixed}\right\}
\end{equation}
This principle determines the most probable distribution of particles at statistical equilibrium --i.e., a macrostate-- that best represents the array of particles at a microscopic level. From a dynamical point of view, the states of these systems towards equilibrium can result from a violent relaxation process, following Lynden-Bell \citep{lynden1967statistical}. When studying the kinetic theory of self-gravitating fermions accounting for tidal truncation (necessary to avoid configurations with infinite mass), one can derive the truncated Fermi-Dirac-like DF given by
\begin{equation}\label{eq:DF}
    f(\Egr) =\frac{2}{h^3}
        \frac{1-\exp{[(\Egr - \Egr_c)/(\sqrt{g_{00}(r)}\beta(r))]}}{1+\exp{[(\Egr/\sqrt{g_{00}(r)} - \alpha(r))/\beta(r)}]},
\end{equation}
A rigorous derivation and further detail can be found in \cite{chavanis1998degenerate,2020EPJP..135..290C}, together with a recent stability analysis applied to the Milky Way in \cite{krut2025thermodynamics} for these fermionic halos. Expression (\ref{eq:DF}) is equivalent to the DF adopted in previous works for self-gravitating fermion configurations in GR \citep[e.g.][]{2018PDU....21...82A,2019PDU....24..278A,2020EPJP..135..290C,2020A&A...641A..34B,2021MNRAS.502.4227A}. In the present case, we have written the DF in terms of the conserved (global) particle energy $\Egr$, rather than the usually adopted local particle energy, $\varepsilon=\sqrt{m^2 + p^2}/m$. These relate by $\Egr = \sqrt{g_{00}(r)}\varepsilon(r)$.
$h$ is the Planck constant, and $\beta(r)$, $\alpha(r)$ are the temperature and relativistic chemical potential per-unit rest mass $m$, i.e.,
\begin{equation}
    \beta(r)= \frac{k_B T(r)}{m},\quad \alpha(r) = \frac{\mu(r)}{m}.
    \label{eq:params}
\end{equation}
The cut-off energy $\Egr_c$ sets a limit for tidal truncation (i.e., $f(\Egr>\Egr_c)=0$). 
It is also convenient to use the fermion degeneracy parameter $\theta$, which corresponds to 
\begin{equation}
      \theta(r)=\frac{\mu(r) -m}{k_B T(r)}  = \frac{\alpha(r)-1}{\beta(r)}.
\end{equation}
%


Here, the temporal metric component $g_{00}(r)$ is that generated by the regular fermionic distribution which satisfies Einstein’s field equations. The associated stress-energy tensor is modeled as a perfect fluid \citep{PhysRev.187.1767}:
\begin{equation}
    T_{\mu\nu} = (\rho + P) u_\alpha u_\beta - P g_{\alpha \beta},
\end{equation}
where $u^\alpha$ is the fluid four-velocity, which obeys $u^\alpha u_\alpha = 1$,  $g_{\alpha \beta}$ is the spacetime metric, and the energy density and pressure are computed according to statistical kinetic theory, by integrating the DF over energy space
\begin{subequations} 
    \begin{align}
    \label{eq:rho-rar}
    \rho(r) &= 4\pi m^4 \int_1^{\varepsilon_c} f(\varepsilon)\varepsilon^2 \sqrt{\varepsilon^2-1}\ d\varepsilon \\
    P(r) &= \frac{4\pi m^4}{3} \int_1^{\varepsilon_c} f(\varepsilon)\left( \sqrt{\varepsilon^2-1}\right)^3d\varepsilon \label{eq:P-rar}
\end{align}
\end{subequations}
Note that in these self-gravitating configurations, the equation of state (EoS), for $P_r(m,\varepsilon_c,\beta,\alpha)$ and $\rho_r(m,\varepsilon_c,\beta,\alpha)$, is not postulated, but rather generated once the DF is known.
Consistency is imposed through the Tolman and Klein conditions \citep{1949RvMP...21..531K}, which are relativistic generalizations of the zeroth and first laws of thermodynamics, together with a cut-off condition $\varepsilon_c(r)$ derived from energy conservation along geodesics. Choosing to write the temporal metric potential in terms of a gravitational potential, $g_{00}(r)=e^{\nu(r)}$, the final form of the model equations reads:
\begin{subequations}
    \begin{align}
    \label{eq:a}
    \frac{dM(r)}{dr} &= 4\pi r^2 \rho(r) \\
    \label{eq:b}
    \frac{d\nu(r)}{dr} &= \frac{2}{r^2}(4\pi r^3P+M) \left[1-\frac{2M(r)}{r}\right]^{-1}\\ 
    \label{eq:c}
    \frac{d\beta(r)}{dr} &= -\frac{\beta(r)}{2} \frac{d\nu(r)}{dr}  \\ 
    \label{eq:d}
    \frac{d\theta(r)}{dr} &= -\frac{1}{2\beta(r)} \frac{d\nu(r)}{dr} \\
    \label{eq:e}
    \frac{d\varepsilon_c(r)}{dr} &= -\frac{\varepsilon_c(r)}{2} \frac{d\nu(r)}{dr} 
\end{align}
\end{subequations}
Equations (\ref{eq:a}) and (\ref{eq:b}) correspond to the well-known Tolman-Oppenheimer-Volkoff equations for hydrostatic equilibrium configurations in GR. Equations (\ref{eq:c}) and (\ref{eq:d}) come from the Tolman and Klein conditions, and (\ref{eq:e}) gives the cut-off parameter.
Given a set $(M_0,\nu_0,\beta_0,\theta_0,\varepsilon_{c,0})$ of initial conditions at the center $r=r_0$, the above system of equations can be integrated to match desired boundary conditions.
%
%
The most general solution leads to a DM density radial profile characterized by a nearly homogeneous, degenerate compact core (dominated by degeneracy pressure and highly sensitive to the particle mass), followed by an extended and diluted halo (dominated by thermal pressure). Such \textit{dense core}--\textit{diluted halo} distributions correspond to positive central degeneracies parameter $\theta_0 \gtrsim 10$, constituting the core-halo family of solutions \citep{2015MNRAS.451..622R,2021MNRAS.502.4227A}. 
For a low central degeneracy parameter $\theta_0 \ll -1$, we obtain the relativistic analog of the King model of a diluted Fermi gas with cutoff (resembling the Burkert DM profile as shown in \citep{2023ApJ...945....1K}). 
%
%
Figure \ref{fig:degeneracy} shows two RAR-DM halos obtained with high and low central degeneracy parameters ($\theta_0 = -33$ and $\theta_0 = +39$, dotted lines). The model omits particle self-annihilation, consistent with the obtained temperatures, $T \ll m/k_B$.

\subsection{Baryon-induced collapse: DM cores into SMBHs} \label{subsec:BIC}

As shown in \cite{2023MNRAS.523.2209A}, the core-halo solutions following equilibrium configurations as the central density grows can reach a state of gravitational core-collapse into an SMBH. These critical cores can be well approximated by configurations in the fully degenerate regime, for which the limiting mass is given by the Oppenheimer-Volkoff relation \citep{PhysRev.55.374}
\begin{equation}\label{eq:Mcrit_DM}
M^{(0)}_{\rm crit}\approx 0.38 \frac{m_{\rm Pl}^3}{m^2} = 6.27 \times 10^7 \left(\frac{\text{100 keV}/c^2}{m}  \right)^2 M_{\odot},
\end{equation}
where $m_{\rm Pl} = \sqrt{\hbar c/G} = 2.18 \times 10^{-5}$ g is the Planck mass. 
As detailed in \cite{2024ApJ...961L..10A}, when considering the combination of DM + baryonic matter, the new equilibrium sequences with their corresponding turning points in the $M$-$\rho_c$ plane lead to new values of the critical mass for gravitational collapse (achieved when $\partial M /\partial \rho_c =0$, with $\rho_c$ the central density). This value will depend on the baryonic-to-DM mass fraction $\chi = M_b /M_{\rm dm}$, which may take values between $0 \leq \chi \leq 0.8$. These heavy SMBH seeds can grow to a few $10^9 \Msun$ in less than 1 Gyr accreting at sub-Eddington rates \cite{2023MNRAS.523.2209A,2024ApJ...961L..10A}.

%
%
A possible sequence of events in the baryon-induced collapse framework is as follows: there is a starting core-halo configuration (pure DM, $\chi=0$) on which the baryonic matter starts to fall into its potential well, modifying its equilibrium state. 
The total configuration evolves until it reaches the gravitational collapse condition. An SMBH is formed with mass $M_{\rm BH} = M_{\rm crit}$ within a fraction of a Gyr for a typical baryon fraction. 
For a realistic baryon-induced core collapse to occur --that is, in $O(t \sim$ Gyr)-- there is a minimum DM core mass above which, when accreting baryons, the critical mass is reached. This threshold value is given by
\begin{equation}
    M_{\rm dm }^{\rm (min)} \approx 0.22 M^{(0)}_{\rm crit}
\end{equation}
for which it corresponds a maximum SMBH formation time $t_{\rm BH}^{\rm (max)} = 0.44 \tau$, where the timescale $\tau$ takes the baryon physics into account
\begin{equation}
    \tau \approx 4.21 \left(\frac{\Msun/{\rm pc}^3}{\rho_b}\right) \left(\frac{v_b}{100\ {\rm km/s}}\right)^3 \left(\frac{10^6 \Msun}{M_{\rm dm}}\right)\ {\rm Gyr}
\end{equation}
where $\rho_b$ is the baryon density at the gravitational capture radius $R_{\rm cap}=2GM/v_b^2$, $v_b$ is the baryons speed and $M_{\rm dm}=M(t=0)$ is the initial core mass (pure DM). 
In the same manner described in section \ref{subsec:framework}, the baryonic induction process occurs in the adiabatic regime when a DM core with mass $M_{\rm dm}^{\rm (min)}$ accretes baryons growing (together) to $M_{\rm crit}$. 
There is, however, a relevant difference relative to the SMBH formation from a negligible (adiabatically growing) BH seed. In this scenario, the SMBH formation occurs almost \textit{instantaneously}, i.e., at a free-fall timescale via DM-core collapse.

We apply this approach in section \ref{sec: model 2}. We consider a starting fermionic core-halo distribution of $mc^2 = 300$ keV and $M_{\rm dm} = 1.5\times 10^6 \Msun$. For these fermions the Oppenheimer-Volkoff mass is $M^{(0)}_{\rm crit}=7\times 10^6 \Msun$ ($\chi=0$). Adopting the value $\chi = 0.5$, the corresponding critic mass for collapse, and hence, the resulting SMBH mass is $M_{\rm crit} = M_{\rm BH} = 4 \times 10^6 \Msun$.
There, we show the results on the redistribution of this fermionic DM halo surrounding an SMBH 
that originated from a baryon-induced collapse of a DM core. In appendix \ref{appendix-B}, we develop the rearrangement of DM orbits when the gravitational potential suffers a sudden change from a fermionic compact core to the most compact solution, a Schwarzschild BH.

\begin{figure}[t]
    \centering
    \includegraphics[width=\columnwidth]{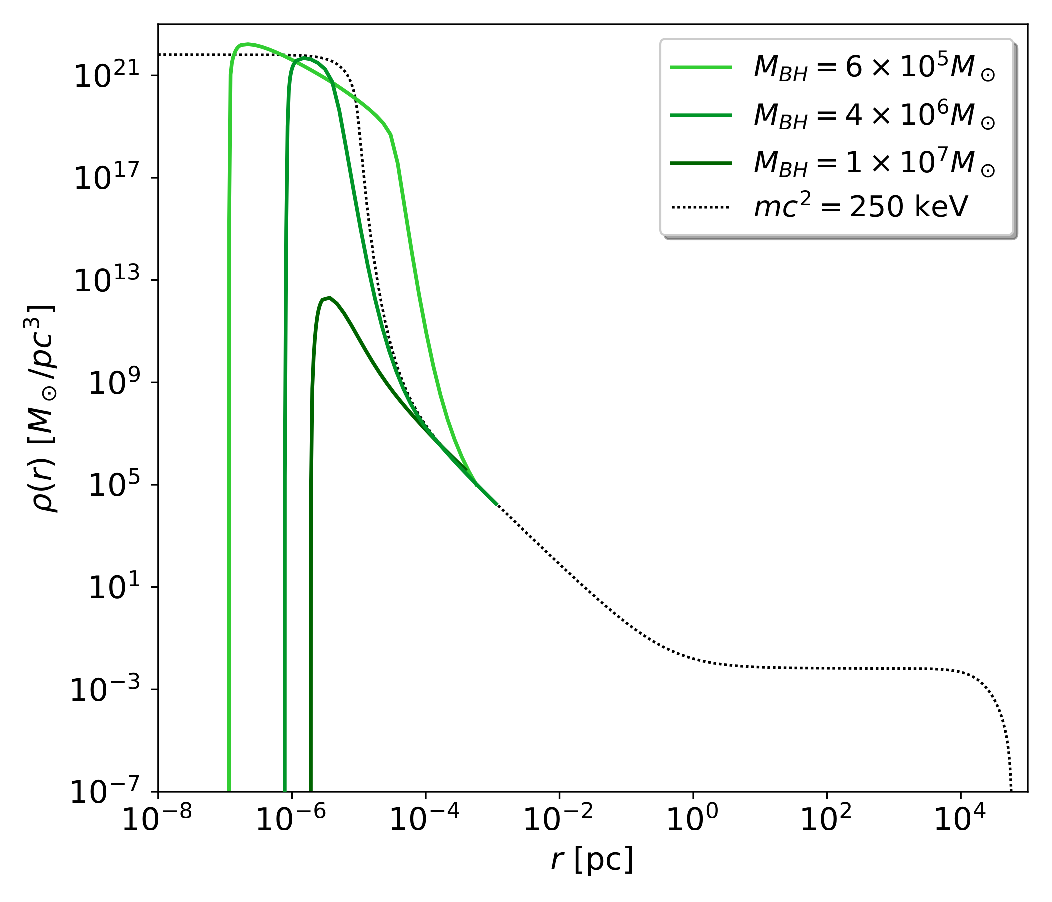}
    \caption{Mass density profiles in green palette, resulting from adiabatic growth of a central BH with final values of $\MBH=6\times10^5,4\times10^6,1\times10^7 \ \Msun$ (corresponding to Model I). The black dotted line corresponds to the original DM halo with a fermion mass of $m=250$ keV.}
    \label{fig:RAR-BHs}
\end{figure}
%
\begin{figure*}[t]
    \centering
    \includegraphics[width=\linewidth]{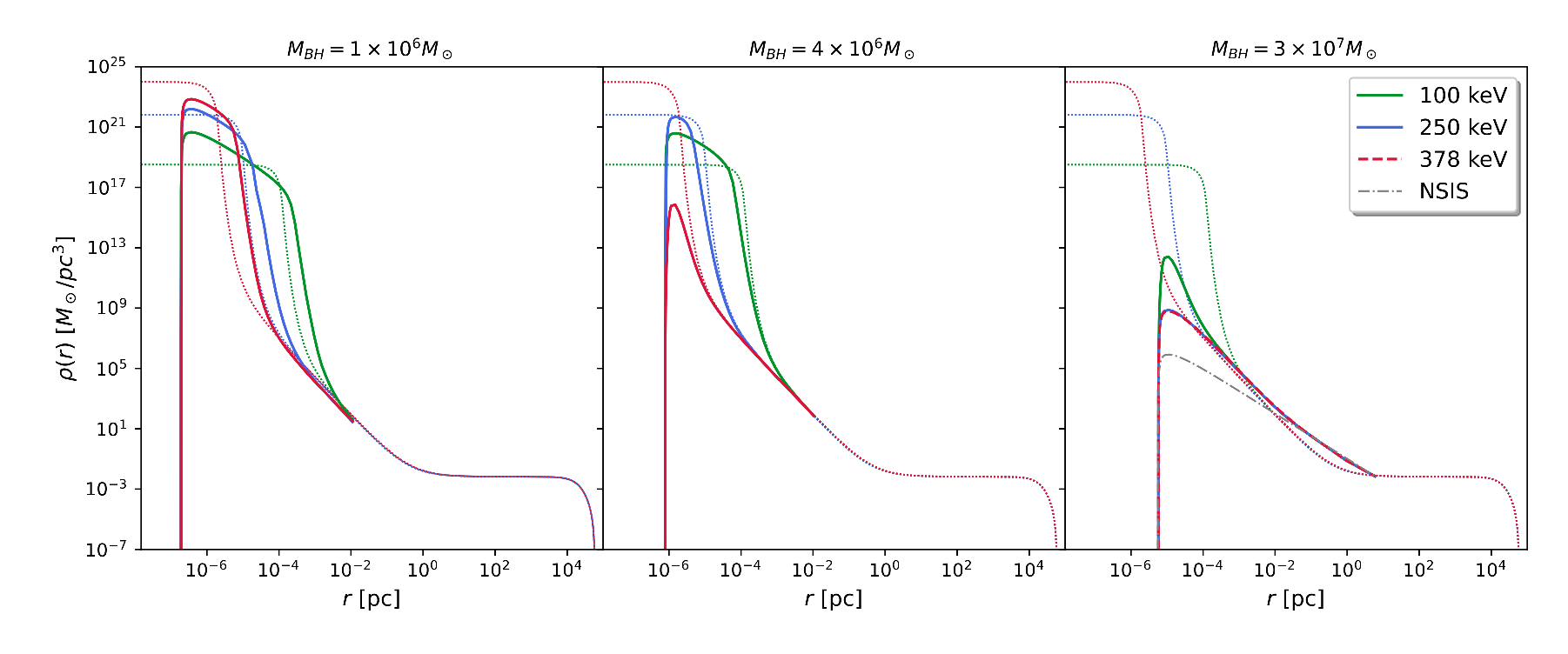}
    \caption{Density profiles corresponding to DM halos with fermion masses of $m=100$, $250$, $378$ keV. The original configurations are represented in dotted lines, while the resulting spikes are in solid lines. As the central BH mass increases, the DM spikes exhibit a change in their behavior. Left: all DM profiles present a power law with $r^{-3/2}$ in its peak. Middle: the spikes get narrower, and the peaks decrease. The size of this depletion depends on the particle mass. Right: for a more massive central BH, all spikes resemble each other, tending to the Boltzmannian spike for a hosting NSIS DM profile.}
    \label{fig:3-BHs}
\end{figure*}
%
\section{Results} 
\label{sec:Results}
\subsection{Model I: BH growth from baryons} \label{sec: model 1}

In this interpretation, we compute the dynamical rearrangement of DM particles, assuming that an SMBH has grown adiabatically in the center within $\sim 10^7$ yr under Eddington-limited accretion.
We implemented different host DM configurations given by the RAR model, for which the same outer halo is shared, i.e., only differing in the central core compactness (see, e.g., \cite{2018PDU....21...82A}). 
In principle, these general results apply to any size and mass of astrophysical halos \cite{2019PDU....24..278A}. We selected those corresponding to typical Milky Way-like galaxies to compare with the previous literature of spikes obtained for different DM halo models.
To achieve this, we imposed appropriate boundary conditions given by $M(r=29 {\rm\ kpc})=1.8\times 10^{11}\Msun$, a total mass of $M(r_b=58 {\rm\ kpc})=2.4\times 10^{11}\Msun$ and a core mass set to $M_c=3.5\times 10^6 \Msun$, as in \citet{2020A&A...641A..34B,2021MNRAS.505L..64B}. 

In Fig. \ref{fig:RAR-BHs}, we present the spike density profiles that arise in a RAR core-halo configuration as the central mass of the SMBH grows. The dotted black curve shows the host halo of $m=250$ keV fermions. 
These profiles could be seen as the evolution of the DM spike as the BH grows. In particular, for this $250$ keV halo, spikes with central BH masses below $\lesssim 4\times 10^6 \Msun$ show an extended power-law behavior following its peak, with index $\gamma= 3/2$. For larger BH masses, the spike becomes sharper and lower.

Figure \ref{fig:3-BHs} compares DM spikes formed within three different halos of particle masses $m=100$, $250$, and $378$ keV, for three selected values of the central BH mass, $10^6 M_\odot$, $4 \times 10^6 M_\odot$, and $3\times 10^7 M_\odot$. 
The same trend of the DM spike evolution holds for all DM particle masses, although the shrinking of the spike highly depends on its value. This behavior is related to the compactness of the original DM core, which increases with the particle mass. Eventually, for sufficiently high BH mass ($\sim 10^7\Msun$), the DM spikes begin to resemble each other, independently of the particle mass, and tend to the spike predicted by the non-singular isothermal sphere (NSIS) profile (Fig. \ref{fig:3-BHs}, right panel). The above occurs because, inside the BH sphere of influence ($\sim 10^6 \Rsch$), the DM particles that have not fallen in (i.e., with $L > 2\sqrt{3}\MBH$) have a low degeneracy parameter, leading to their behavior in a Boltzmannian regime.

\begin{figure}[t]
    \centering
    \includegraphics[width=\columnwidth]{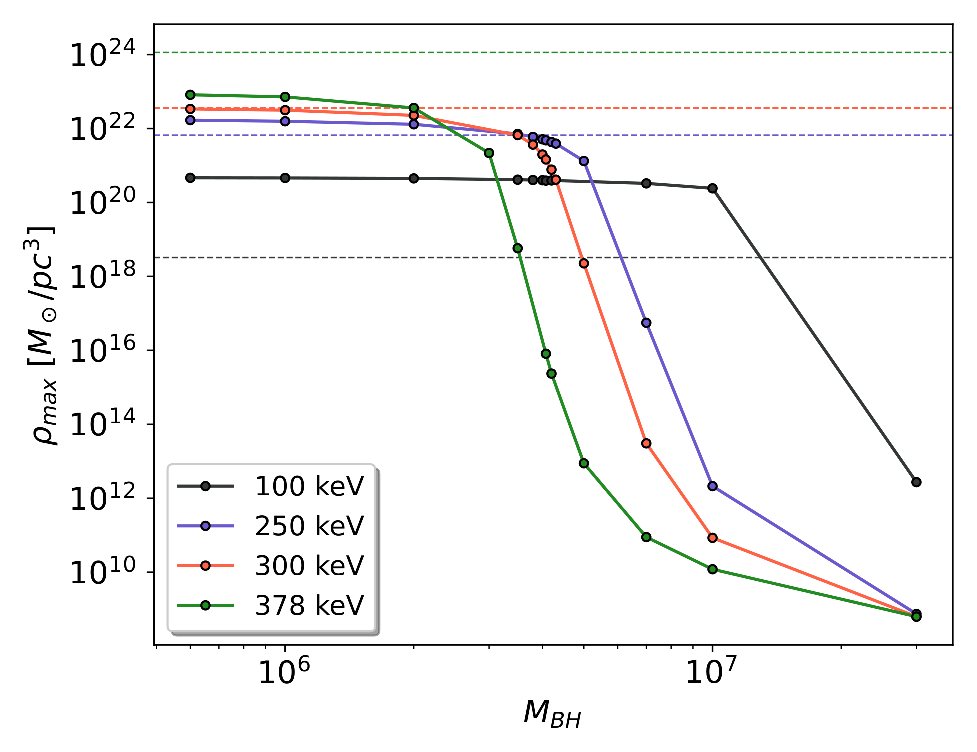}
    \caption{Dependence of the peak heights with the central BH mass, ranging from $\MBH=6\times10^5\Msun$ to $\MBH=3\times10^7\Msun$, for different DM particle masses. The dashed horizontal lines correspond to the values of central densities of the original  M halos. See also Figure \ref{fig:3-BHs} for selected spikes with given central $\MBH$ values.}
    \label{fig:rho_vs_MBH}
\end{figure}

Figure \ref{fig:rho_vs_MBH} shows the tendency of the maximum density $\rho_{\rm max}$ of the peak of the spike as a function of the BH mass for several fermion masses. The dashed horizontal lines correspond to the central density $\rho_0$ of the original DM halos. The shrinkage of the peaks occurs earlier as the particle mass increases. For BH masses where the peaks maintain approximately the same height (e.g., $100$ keV, black line from $\MBH \lesssim 10^7 M_\odot$), is the region of power-law behavior, where the spike profile only changes in its width (see left and middle panels of Fig. \ref{fig:3-BHs}, $100$ keV green lines). Interestingly, at fixed fermion mass, there is a BH mass over which the spike density depletes instead of being enhanced relative to its initial value (i.e., compared with the horizontal dashed lines). Furthermore, there is a threshold particle mass, $\mth \sim 300$ keV, above which there is no DM density enhancement for any BH mass. 

\begin{figure}[!t]
    \centering
    \includegraphics[width=\columnwidth]{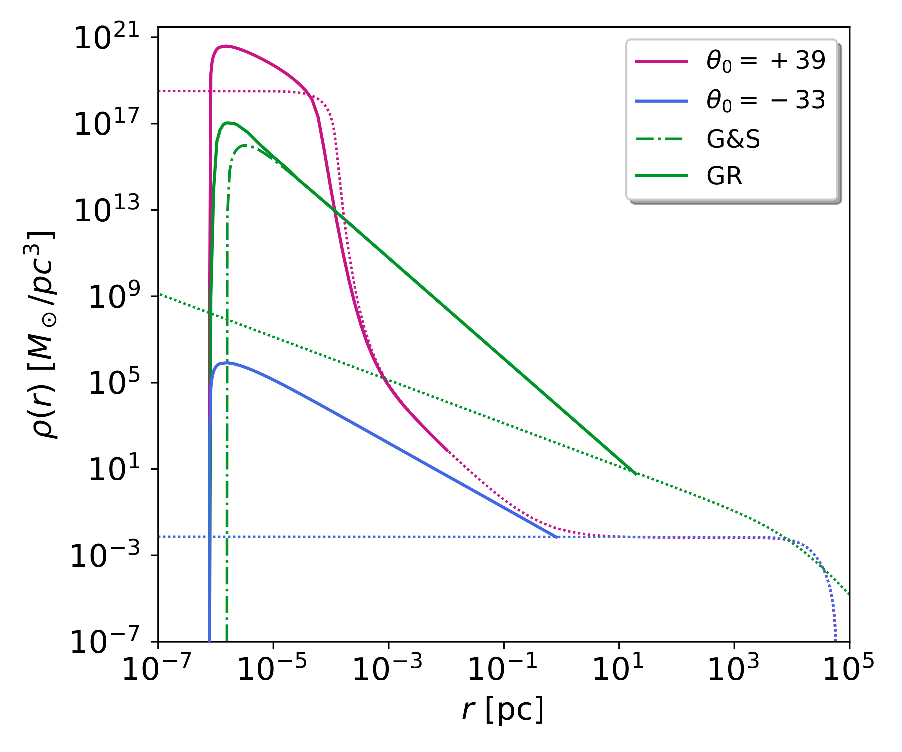}
    \caption{Density profiles corresponding to original DM halos with fermion mass of $m= 100$ keV. In purple for a core-halo solution with a high central degeneracy parameter of $\theta_0=39$, in blue for a low central degeneracy solution with $\theta_0=-33$, corresponding to the Boltzmannian regime. For comparison, the DM spike of G\&S is shown by the dotted-dashed green curve and its relativistic correction in solid green (adapted from \cite{PhysRevD.106.044027}). These spikes are obtained from a typical NFW halo, shown in dotted green. The chosen central BH mass in all cases is $\MBH=4 \times 10^6 \Msun$ to contrast with the literature.}
    \label{fig:degeneracy}
\end{figure}

Figure \ref{fig:degeneracy} compares the sizes and scales of three different DM spikes for the same central BH. We set the value to $\MBH=4\times 10^6 \Msun$ to compare with the spike of G\&S and other works.
For two of the spikes, we adopt the RAR model for $m=100$ keV fermions as the host halo, differing in the central degeneracy parameter value. For $\theta_0=39$, the original profile is of the core-halo family, whereas for $\theta_0=-33$, the profile is that of (relativistic) NSIS with cutoff. 
For the remaining spikes, we adopt as the original DM halo a power-law density model with index $r^{-1}$, corresponding to the inner region of a Navarro-Frenk-White (NFW) DM halo \citep{navarro1997universal}. 
In this case, for the sake of comparison with G\&S, we consider a minimum value for the angular momentum given by $L_c=2\Rsch$, which leads to a spike vanishing at $4\Rsch$ instead of $2\Rsch$.
We also adopt the spike obtained in the fully relativistic treatment computed in \cite{PhysRevD.106.044027}.

It is worth noticing that, for RAR configurations with low central degeneracies (where the fermions behave in a Boltzmannian regime), we recover the spike found by G\&S and other authors, where the density goes with $r^{-3/2}$, as expected for models with finite cored-halos (Fig. \ref{fig:degeneracy}, blue lines). 
A first observation to be drawn from these results is that although higher spike peaks are found with RAR host halos, the DM overdensity is always confined to a narrower region compared to spikes arising from power-law profiles, which are lower and more widespread.
%

\subsection{Model II: DM core collapse} \label{sec: model 2}

To handle this scenario, we combine the two formalisms detailed in sections \ref{subsec:framework} and \ref{subsec:BIC}. We separate the problem into two stages:\\ 
\indent i) We start with an initial (sub-critical) core-halo equilibrium configuration. The DM core, of mass $M_{\rm dm} \geq M_{\rm dm}^{(\rm min)}$, accretes baryons and consequently grows to its critical mass for gravitational collapse, $M_{\rm crit}$. The timescale on which this occurs is $\sim 0.1$--$1$ Gyr \cite{2024ApJ...961L..10A}. Therefore, the core grows in the adiabatic regime. \\
\indent ii) Once the compact core has reached the critical mass, the gravitational collapse of the core sets in. All the particles belonging to the core collapse into an SMBH. The remaining particles orbit in a new, suddenly changed gravitational potential.

\begin{figure}[!t]
\centering
    \includegraphics[width=\linewidth]{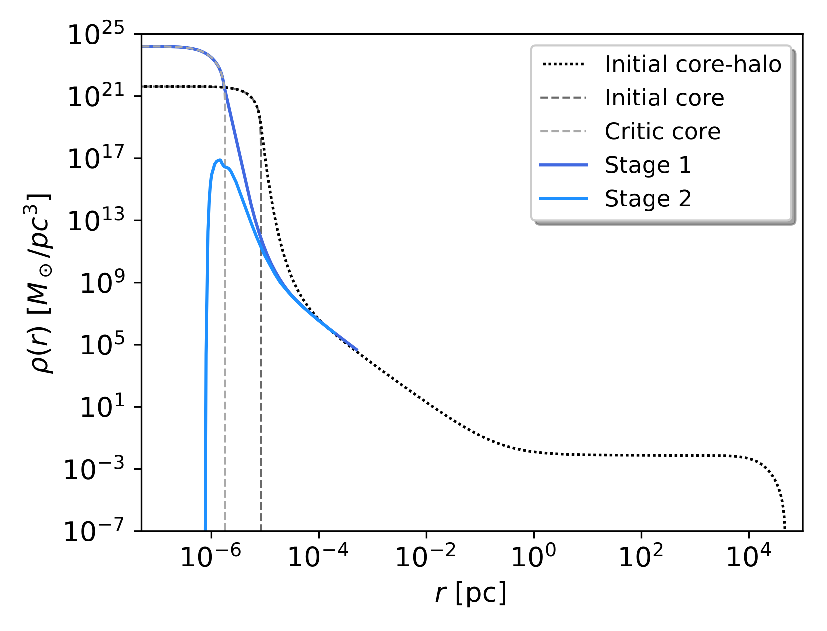}
    \caption{Evolutionary process of a core-halo DM configuration of 300 keV fermions (dotted black).
    In stage 1, the central core (initial core, dashed gray) adiabatically grows as it accretes baryons until it reaches the critical mass for gravitational collapse (critical core, dashed-dotted gray). Consequently, the particles re-accommodate under the core central potential (solid dark blue). The critic values are $M_{\rm crit}=4\times 10^6 \Msun$ and a baryon to DM mass fraction of $\chi=0.5$. After collapse is triggered, an SMBH is formed with $\MBH=M_{\rm crit}$, and the remaining DM forms a spike density shown in solid blue.}
    \label{fig:model2}
\end{figure}
\begin{figure}[!t]
    \centering
    \includegraphics[width=\linewidth]{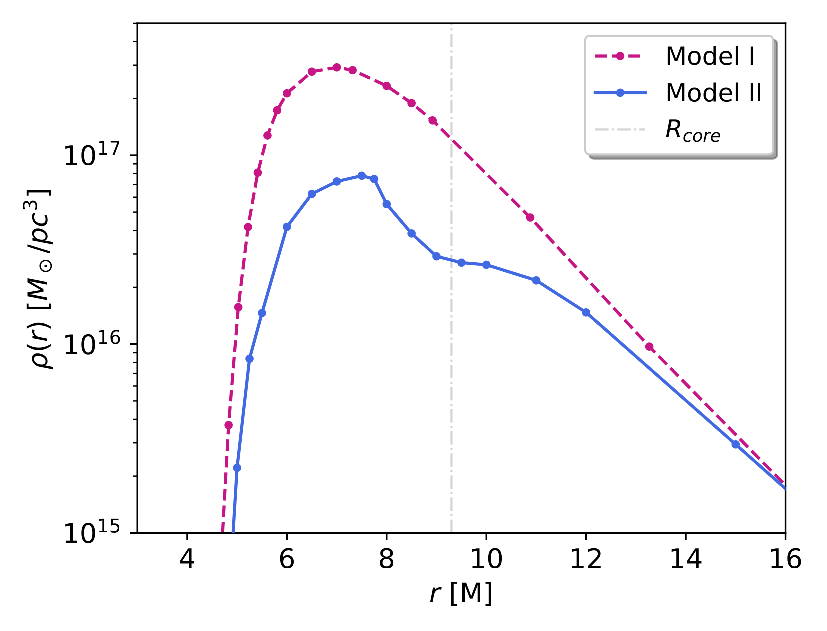}
    \caption{Comparison of spike profiles for the same central $\MBH=4\times 10^6 M_\odot$, formed in a core-halo DM configuration of 300 keV fermions. In Model I, the DM particles rearrange around an adiabatically growing BH (dashed purple). In Model II, the remaining DM particles form the density profile of the spike after the core collapse (solid blue). The critical core radius is $R_{\rm core} = 9.3 M_{\rm crit}$.}
    \label{fig:model1and2}
\end{figure}

To illustrate this method, we selected an initial equilibrium configuration for $mc^2=300$ keV fermions, with a core of mass $M_{\rm dm}=1.5\times 10^6 \Msun$.
The baryonic to DM mass fraction at the time of collapse is $\chi = 0.5$, with corresponding critic mass $M_{\rm crit} =4\times 10^6 \Msun$ according to the baryon-induced collapse scenario \citep{2024ApJ...961L..10A}. Instead, the pure DM critic core ($\chi=0$) corresponds $M^{(0)}_{\rm crit}=7\times 10^6 \Msun$. In Fig. \ref{fig:model2}, we show the evolutionary process, according to model II, of the core-halo DM profile.
For the first stage, we perform the numeric computation of the new DM density using equation \ref{eq:rho}, where $g_{00}(r)$ now corresponds to the metric produced by the compact critic core $M_{\rm crit}$ (see Fig. \ref{fig:metrics}), and the integration limits in $\Egr$ and $L$ correspond to a regular distribution. 
As expected, we found the redistribution of DM particles closely resembling that of a core-halo profile in which the core is critical (see Fig. \ref{fig:model2}, stage 1). When the core undergoes a (radial) gravitational collapse, the remaining DM particles change their energy, leading to orbits with inner pericenters, which are limited by $2\Rsch$ (see Fig. \ref{fig:model2}, stage 2). 
In Fig. \ref{fig:model1and2}, we compare the resulting spikes given by models I and II, for the same central BH mass $\MBH$ and core-halo host DM profile.
We find these results with a slight difference, showing up only at the peak.
The flattening of the peak in Model II corresponds to the subtraction of DM particles that belonged to the core at the time of its collapse. The details of this numeric computation are shown in Appendix \ref{appendix-B}.



\section{Discussion and Conclusions}  \label{sec:conclusions}

Much is being discussed in the literature on the theoretical prediction that DM around BHs enhances its density, forming the so-called DM spikes. Specifically, it has been widely assumed and adopted that the density profile of such DM spikes obeys a power-law behavior. In this work, we have calculated the density spikes of DM made of massive, neutral fermions within a general relativistic analysis. We have obtained the initial DM profiles within the RAR model framework, which adopts the fermion gas at finite temperatures in hydrostatic and thermodynamic equilibrium (see section \ref{subsec:RAR model}). We assess different states of central degeneracy of the fermions and different particle masses. 
We have adopted two possible interpretations for the BH-seed origin and evolution, Models I and II. In the former, the spikes follow from a relativistic extension of the G\&S method detailed in Section \ref{subsec:framework} and Appendix \ref{appendix-A}. In this framework, the spikes are formed and evolve adiabatically within a self-gravitating DM halo in the presence of a negligible BH seed (possibly of ordinary baryonic origin).
The key results are:
\begin{enumerate}
    \item Given a central $\MBH$, the fermionic spike density profile crucially depends on the mass of the DM particles and their physical regime (state of degeneracy). This is markedly different from the spike density profile from power-law DM density halos, which depends only on the initial power-law index $\gamma$. 
    \item The DM spikes arising from fermionic core-halo configurations do not develop a simple power-law profile (see, e.g., Figs. \ref{fig:RAR-BHs} and \ref{fig:3-BHs}).
    \item For a given fermion mass, there is a specific mass of the central BH above which there is no density enhancement. Over such a mass, the DM density decreases relative to the initial one (see Fig. \ref{fig:rho_vs_MBH}). For fermion masses above $\sim 300$ keV, there is no enhancement of the density of the DM surrounding the BH for any BH mass, but a suppression.
    \item The typically assumed power-law DM spikes occur only in the case of initial DM configurations in the Boltzmannian regime. In the RAR case, when the central fermions are in a state of low degeneracy ($\theta_0 \ll -1$), we recover the resulting DM spike of $\rho \propto r^{-3/2}$ typical of cored halos as the NSIS (see Fig. \ref{fig:degeneracy}). 
\end{enumerate}%
If one could constrain the DM distribution at $r \lesssim 1$ pc scales, given different astrophysical observables, these results would imply an indirect way to predict whether RAR-DM halos are born in a core-halo regime ($\theta_0 \gg 1$), or only as a diluted halo ($\theta_0 \ll -1$).

It is worth noting that this theoretical treatment has some limitations and approximations.
In particular, the G\&S method, strictly speaking, requires that in the initial state, the DM halo is self-gravitating. However, once the BH has grown, the gravitational influence of the DM particles is purely due to the BH, and the gravitational potential of the remaining DM mass is neglected. For RAR-DM halos, the enclosed mass inside the grown BH sphere of influence ($\sim 10^6 \Rsch$) will depend on the fermion mass and not only on the central BH mass. For fermions of $mc^2 \lesssim 250$ keV and $\MBH \lesssim 4\times  10^6 M_\odot$, the enclosed DM mass is comparable to that of the central BH, $M_{\rm DM} \sim \MBH$. This fact favors the RAR spike profiles with fermion masses $\gtrsim 300$ keV. Further requirements are that the DM halo obeys spherical symmetry and the BH seed cannot be off-center \cite{ullio2001dark,merritt2003single}.

We turn to Model II. It encompasses a new self-consistent interpretation of the problem of how SMBHs form and evolve at the center of DM halos from a cosmological viewpoint. In this case, violent relaxation predicts that DM fermions can describe spherical distributions in thermodynamic equilibrium, where the most general morphology is a degenerate compact core surrounded by a diluted halo \citep{chavanis1998degenerate,chavanis2015models,2021MNRAS.502.4227A}. These core-halo DM profiles are stable (local maxima of entropy) and extremely long-lived. There is a critical point of instability where core collapse towards an SMBH occurs. This gravitational collapse can only occur for halo virial masses of $M_{\rm vir} \gtrsim 10^9 \Msun$, starting at early-universe stages $z_{\rm vir} \approx 10$ \citep{2021MNRAS.502.4227A}. Furthermore, for DM cores that are not in a critical state (i.e. $\partial M_{\rm c}/\partial \rho_{\rm c}=0$), the collapse can still be triggered by baryonic accretion in a fraction of a Hubble time, producing heavy seeds of DM origin for SMBHs at galactic centers \citep{2024ApJ...961L..10A}. In this approach, we calculated the redistribution of the remaining fermionic DM particles after the collapse of the critical core. The same key findings (1, 2, and 3 in this discussion) also apply to this method. However, one difference in point 1 is that the morphology of the fermionic DM spike now depends on the value of the fraction of baryons inside the DM core $\chi$, in addition to the mass of the fermion and its state of degeneracy (for a given central $\MBH$).

The current theoretical framework can be applied to various DM astrophysical probes. In particular, the enhancement of the DM density can have a relevant impact, among others, on stellar dynamics (e.g., stars orbiting BHs), the rate at which a BH grows by accreting DM, and the dynamical friction of DM particles on the motion of BHs in binaries. The latter effect can be relevant in the so-called \textit{final parsec problem} of merging SMBHs. Here, we have focused on the DM spikes around BHs of large masses. It remains an interesting topic to assess whether or not DM also forms spikes around stellar-mass BHs or compact stars, e.g., neutron stars, given their shorter lifetimes. This topic, which has hardly been addressed, is especially relevant for analyzing the possible role of DM in low-mass and high-mass X-ray binaries \citep{chan2023indirect,qin2024alternative}.

\vspace{-0.5cm}
\begin{acknowledgments}
This work used computational resources from CCAD $-$ Universidad Nacional de Córdoba (\href{https://ccad.unc.edu.ar/}{https://ccad.unc.edu.ar/}), which are part of SNCAD - MinCyT, República Argentina. \\
V. Crespi thanks the financial support from CONICET, Argentina. \\
C. R. A. thanks the financial support from CONICET, Argentina, and ICRANet, Italy.
\end{acknowledgments}

\newpage
\appendix
\section{Energy relation} \label{appendix-A}
Consider a spherically symmetric spacetime with the line element 
\begin{equation}
    ds^2 = g_{00}(r)dt^2+g_{11}(r)dr^2-r^2 d\theta^2 - r^2 \sin\theta d\phi^2.
\end{equation}

A massive, free-falling test particle in this geometry possesses four Killing vectors that lead to conserved quantities, namely the energy per unit rest mass $\Egr$, and the three components of the angular momentum per unit rest mass $L_x$, $L_y$, and $L_z$. 

In the space of angle-action variables, the actions are
\begin{align}
    I_\phi(L_z) &= \frac{1}{2\pi} \oint L_z\ d\phi, \label{eq:action_phi} \\
    I_\theta(L,L_z) &= \frac{1}{2\pi} \oint \sqrt{L^2 - \frac{L_z^2}{{\rm sin}^2(\theta)}}\ d\theta, \label{eq:action_theta} \\
    I_r(\Egr,L) &= \frac{1}{2\pi} \oint \dot{r} \ dr, \label{eq:action_r}
\end{align}
where the radial component of the four-velocity is 
\begin{equation}
    \dot{r}^2 = \frac{\Veff^2(r,L)-\Egr^2}{g_{00}(r)g_{11}(r)},
\end{equation}
and the form of the effective potential is 
\begin{equation}
    \Veff^2(r,L)=g_{00}(r)\left(1+\frac{L^2}{r^2}\right).
    \label{eq:potential}
\end{equation}
In the case of a Schwarzschild BH, we have $g_{00}(r)g_{11}(r)=-1$. The azimuthal action \ref{eq:action_phi} becomes trivially $I_\phi = L_z$, the latitudinal action \ref{eq:action_theta} is $I_\theta = L - L_z$ and the radial action becomes
\begin{equation}
    I_r(\Egr,L)=\frac{1}{\pi} \int_{r_p}^{r_a} \sqrt{\frac{\Veff^2(r,L)-\Egr^2}{g_{00}(r)g_{11}(r)}} \ dr
\end{equation}
where $r_p$ and $r_a$ are the pericenter and apocenter of the orbit, respectively. Suppose the central potential that confines the particles varies \textit{slowly}, i.e., when the SMBH grows on a much longer timescale than the dynamical timescale. In that case, the actions can be assumed constant in time, constituting the adiabatic invariants \cite{binney2011galactic}. Due to their conservation, the DF of the system is also conserved between an initial and a final state
\begin{equation}
    f_i(\Egr_i,L_i)=f_f(\Egr_f,L_f).
\end{equation} 

Equations (\ref{eq:action_phi}) and (\ref{eq:action_theta}) imply the conservation of angular momentum $L_i = L_f$, and the invariance of the radial action allows us to numerically relate the initial and final energies as $\Egr_i = \Egr_i(\Egr_f,L)$. 

\begin{figure}
    \centering
    \includegraphics[width=\columnwidth]{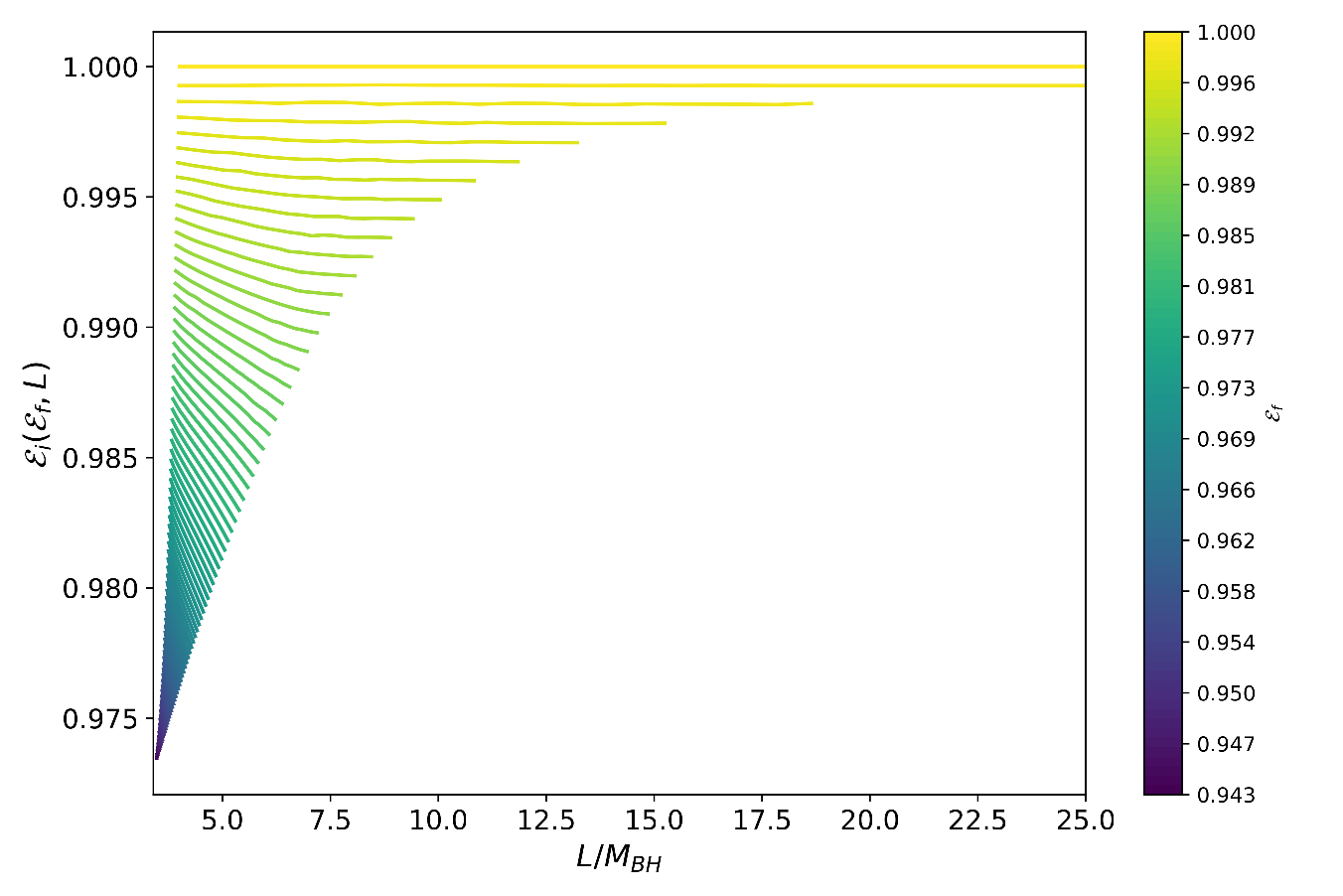}
    \caption{Color map of the surface of original energies $\Egr_i$ for DM particles of $m=250$ keV, as a function of final energies $\Egr_f$ and angular momentum $L$ for particles orbiting a BH of mass $\MBH=4 \times10^6 \Msun$. The color bar corresponds to final energies $\Egr_f$ of bound orbits, permitted from $\Egr_{f,min}=\sqrt{8/9}$ to $\Egr_{f,\rm max}=1$.}
    \label{fig:RA-surface}
\end{figure}

We first compute the turning points $r_p$ and $r_a$ as a function of $\Egr$ and $L$, achieved for the radii where $\Egr=\Veff(\Egr,L)$ using the python SciPy \citep{2020SciPy-NMeth}
function \texttt{scipy.optimize.brenth}; 
then we calculate the radial action for every $\Egr$, as a function of $L$, $I_r(L)$ integrating with \texttt{scipy.integrate.quad};  and look for the intersections when 
\begin{equation}
    I_{r,\Egr_i}(L) = I_{r,\Egr_f}(L)
\end{equation}
This hypersurface in $(\Egr_i,\Egr_f,L)$ space is shown in Fig. \ref{fig:RA-surface} for the case of an initial DM core-halo configuration for fermions of rest mass $m = 250$ keV, core mass of $M_{\rm c} = 3.5\times 10^6 \Msun$ and central degeneracy of $\theta_0=41$.

The range in original energy is  $\Egr_i=[\sqrt{g_{00}(r_0)},1]$, where $g_{00}(r)$ corresponds to the fermionic-only metric tensor and $r_0=0$; the range in final energy is  $\Egr_f=[\sqrt{8/9},1]$, where $\sqrt{8/9}$ corresponds to the lowest possible energy achieved by $L=L_{\rm cr}$ (under the Schwarzschild metric); and the range in angular momentum for every energy is delimited by $L=[L_{-}(\Egr),L_{+}(\Egr)]$, where
\begin{equation}
    L_{\pm}^2(\Egr)= \frac{27\Egr^4\ -\ 36\Egr^2\ +\ 8\ \mp\ \Egr(9\Egr^2-8)^{3/2}}{2(\Egr^2-1)}\ \MBH.
\end{equation}
These functions correspond to $\Veff(r_{\pm},L_{\pm})=\Egr$ when the energy matches the minimum and maximum of the potential (for $r_+,L_+$ and $r_-,L_-$ respectively), giving the characteristic shape in Fig. \ref{fig:RA-surface}.

Finally, to compute the DM density resulting from the adiabatic growth of the central SMBH [Eq. (\ref{eq:rho})], we interpolate the hypersurface using \texttt{scipy.interpolate.LinearNDInterpolator} for any possible value of $\Egr_f$ and $L$ and perform the double integral with \texttt{scipy.integrate.dblquad}.

\begin{figure}
    \centering
    \includegraphics[width=\columnwidth]{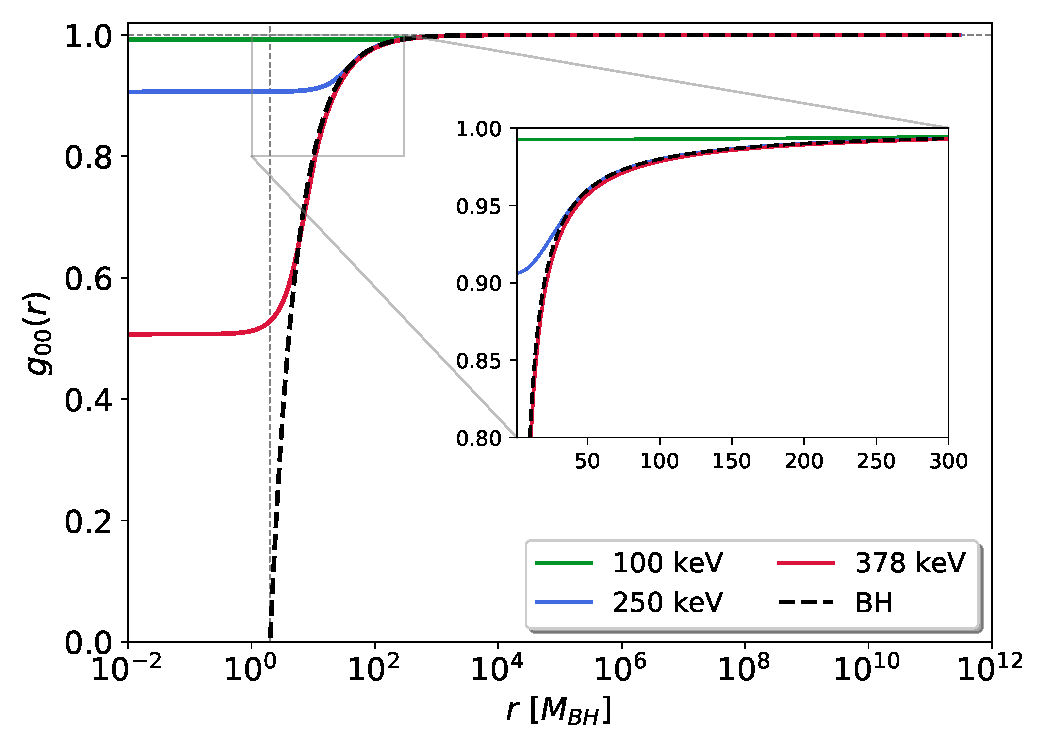}
    \caption{Full extent of the gravitational metric potential of fermionic core-halo configurations. Solid lines correspond to the same solutions shown in Figure \ref{fig:3-BHs} (with corresponding colors). The black-dashed line corresponds to the Schwarzschild BH metric solution given in Eq. (\ref{eq:g00}).}
    \label{fig:g00-full}
\end{figure}

\newpage
\section{Instantaneous BH collapse} \label{appendix-B}

\begin{figure}[t]
    \centering
    \includegraphics[width=\columnwidth]{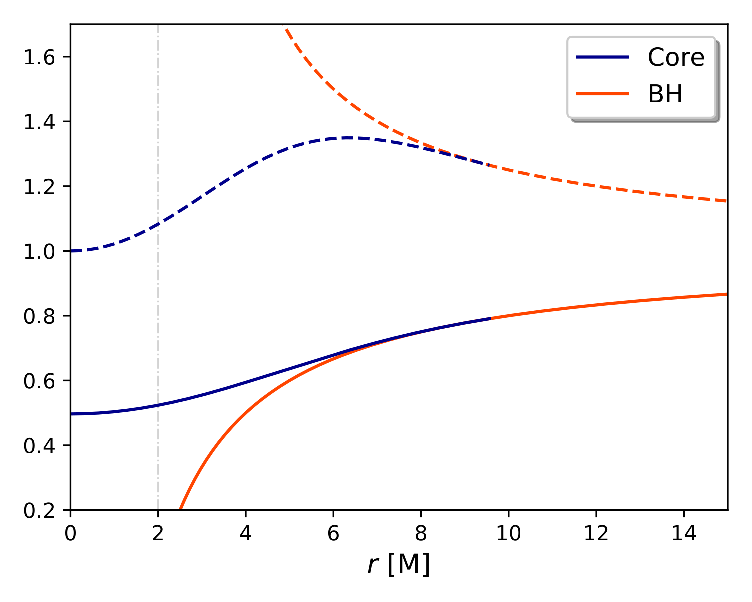}
    \caption{Metric potentials for the DM critical core in blue and a Schwarzschild BH in red, computed for the same mass $M_{\rm crit}=\MBH$. Solid lines correspond to $g_{00}(r)$ and dashed lines to $-g_{11}(r)$. The core metric smoothly matches the BH at the border, $R_{\rm crit}=9.3\ M_{\rm crit}$.}
    \label{fig:metrics}
\end{figure}

We compute the new orbits of DM particles when the gravitational potential suddenly changes as the DM core collapses to an SMBH. This time interval is considered instantaneous compared to the dynamical timescales. 
We adopt the methodology assumed by \cite{ullio2001dark}, where the DM particle conserves its angular momentum $L$ since the collapse is radial. For a particle located at $\mathbf{r}_i$ with velocity $\mathbf{v}_i$ under the influence of $\Veff$ [Eq. (\ref{eq:potential})] before core collapse, its velocity remains unchanged $\mathbf{v}_i(r_i) = \mathbf{v}_f(r_i)$ right after collapse when the potential corresponds to Eq. (\ref{eq:potential}) with $g_{00}(r)=1-2\MBH/r$. In Fig. \ref{fig:metrics}, we show the difference in the metric potentials between the regular critical core and a Schwarzschild BH solution. 
Since we want to perform the calculation for a general kind of (bound) Keplerian orbit, it is convenient to use the particle energy and angular momentum, as done in \cite{2024arXiv240408731B}.

The probability to find a particle at radius $r_f+dr_f$, that before core collapse was at radius $r_i$ with velocity $\mathbf{v}_i(r_i)$ is 
\begin{equation}
    P(r_f|\Egr_i,L) dr_f= \frac{2d\tau}{T(\Egr_i,L)}=\frac{2}{T(\Egr_i,L)}\frac{dr_f}{\sqrt{\Egr_f^2 - \Veff^2(r_f,L)}},
    \label{eq:probability}
\end{equation}
where $T$ is the orbital period. The change in energy is given by the condition $\mathbf{v}_i(r_i) = \mathbf{v}_f(r_i)$, that in general relativity, for the radial component becomes 
\begin{equation}
    \Egr_f^2 = \Egr_i^2+ \mathcal{V}_{{\rm eff},f}^2(r_i,L)- \mathcal{V}_{{\rm eff},i}^2(r_i,L).  
\end{equation}
\\

The final distribution density is thus obtained by integrating the initial density, weighted by this probability, over the appropriate range of radius $[r_1,r_2]$ that can contribute to the particle being located at radius $r_f$ after collapse
\begin{equation}
    \rho(r_f)=\frac{1}{r_f^2} \int_{r_1}^{r_2} dr_i\ r_i^2\rho_i(r_i)\ P(r_f|r_i,\mathbf{v}_i).
\end{equation}
To find $[r_1,r_2]$ for every $r_f$, we need the explicit dependence on $\Egr_i$ and $L$. This results in computing the triple integral
\begin{equation}
    \rho(r_f)=\frac{4\pi m^4}{r_f^2}\int dL d\mathcal{E}_i dr_i \frac{L\ \mathcal{E}_i\ P(r_f|\Egr_i,L)\ f_i(\mathcal{E}_i,L)}{\left[g_{00}(r_i)(\mathcal{E}_i^2 - \mathcal{V}_i^2(r_i,L))\right]^{1/2}}.
    \label{eq:rho_IBHC}
\end{equation}
Additionally, we require that no particle inside the core at the moment of collapse can contribute to the final density, as it will become part of the SMBH. We impose $r_1,r_2 \geq R_c$. For the case studied in this work, the critical core of $300$ keV fermions has a mass and radius of $M_{\rm crit}=4\times 10^6 \Msun$, $R_{\rm crit}=9.3M_{\rm crit}$.
We compute equation \ref{eq:rho_IBHC} using \texttt{scipy.integrate.tplquad}.

\begin{figure}[t]
    \centering
    \includegraphics[width=\columnwidth]{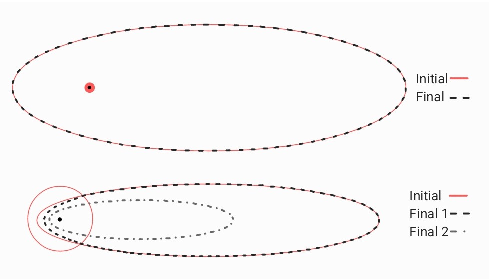}
    \caption{Illustration of the orbital change when the critical core collapses to an SMBH. There are two cases for the initial state of the DM particle. Top: the initial DM particle orbits entirely outside the critical core (solid red). When the collapse occurs, its energy, hence its orbit, remains unchanged (dashed-black). The gravitational potentials are identical outside the core radius (see Fig. \ref{fig:metrics}). Bottom: the initial DM particle has its pericenter inside the core (the solid red circle is the core, and the solid red ellipse the particle's orbit). Two possible endings result in this case. If the DM particle was outside the core at the collapse, its orbit changes slightly by reducing its pericenter (Final 1, dashed black). If the DM particle is inside the core at the collapse, its orbit changes completely (Final 2, dotted dashed gray). However, the particle now belongs to the SMBH, so we subtract its contribution to the final DM density profile.}
    \label{fig:orbits}
\end{figure}
%

\bibliography{references}

\end{document}